# Student Behavior and Epistemological Framing: Examples from Collaborative Active-Learning Activities in Physics

*Rachel E. Scherr and David Hammer*
*Physics Education Research Group, Department of Physics*
*University of Maryland, College Park, MD 20742*

Questions of participant understanding of the nature of an activity have been addressed in anthropology and sociolinguistics with the concepts of *frames* and *framing*. For example, a student may frame a learning activity as an opportunity for sensemaking or as an assignment to fill out a worksheet. The student's understanding of the nature of the activity affects what she notices, what knowledge she accesses, and how she thinks to act. Previous analyses have found evidence of framing primarily in linguistic markers associated with speech acts. In this paper, we show that there is useful evidence of framing in easily observed features of students' behavior. We apply this observational methodology to explore dynamics among behavior, framing, and the conceptual substance of student reasoning in the context of collaborative active-learning activities in an introductory university physics course.

## I. INTRODUCTION

Large college physics courses are especially challenging settings for physics instruction. Among the challenges is the difficulty of providing opportunities for student discussion of physics concepts. For about twenty years, reform-minded physics professors have been using "tutorials" (McDermott, Shaffer, & the Physics Education Group at the University of Washington, 1998), worksheets that guide collaborative active learning in small groups around core conceptual issues identified by research as essential for students to address. Tutorials can be implemented as a relatively small modification of the conventional lecture-based course, and there is now a substantial body of evidence that they lead to substantial improvement in students' conceptual understanding (Redish & McDermott, 1999).

That evidence, however, is almost exclusively in the form of pre- and post-testing with qualitative and quantitative examinations. While gains in student scores support researchers' conjectures regarding what contributes to student learning, there has been little direct investigation of what happens during the instructional activities. Meanwhile, these curricula have not led to corresponding gains in scores on measures of student expectations and epistemologies — how students understand what knowledge, reasoning, and learning in physics entail. In fact, students in most introductory courses, including those that use tutorials, tend to come away with evidently less sophisticated views of knowledge and learning in physics courses (Redish, Saul, & Steinberg, 1998). This seems discrepant: Shouldn't students who are apparently more successful at learning also become more aware of what successful learning entails? Why would experiences that improve conceptual understanding lead to "deterioration" (Redish et al., 1998) of expectations?

There are several possibilities, some of which give cause for concern. One is that students come to see introductory physics as detached from everyday experience: surveys at the end of every course in Redish et al.'s (1998) study indicated students saw physics as less connected to their experience than they did at the start. Another possibility is that student epistemologies as measured by an explicit survey do not reflect their epistemologies in contexts of reasoning within the course; responding to a survey is a very different context from reasoning about physical phenomena (Hammer, 1994). Perhaps students' "practical epistemologies" (Sandoval, 2005) come to be more sophisticated than the survey detects.

The possibly problematic relationship between students' conceptual and epistemological learning motivates us to study students' work during tutorials in detail. In particular, we hope to learn how they understand the activity of tutorials with respect to knowledge, reasoning, and learning. What are their perceptions of its goals, and what methods do they believe will serve them best? In other words, as we elaborate below, we are interested to study students' *epistemological framing* of the activity. To address the possibility of contextual dependence, we study their work *in situ*, rather than interrupting it with a survey or specialized interview.

At the start of this study, the first author set out to formulate a systematic observational protocol: what could we look for in the video data as evidence of how students understand the activity? Initial exploration of video data led to the discovery of four distinct patterns of the students' physical behavior. In some segments of video, for example, the students are bent over their worksheets, talking in subdued voices with their hands relatively still; in other segments, they are sitting up straight, looking at each other, speaking in loud voices, and gesturing prolifically. These clusters of co-occurring behaviors stand out clearly; having had them described, and with very little training, independent coders achieve 90% agreement in classifying 5-second video segments into one of four clusters.

Our purpose in this paper is to argue that these different behavioral clusters are evidence of — and in dynamic interaction with — student epistemologies. Roughly, we will show, the behavioral cluster of sitting up, speaking clearly and gesturing supports and is supported by a framing of the activity as discussing each others' conceptual ideas. In this way, we hope to contribute to research methodology for the study of student learning, and specifically students' epistemological framing. As well, we discuss how this work furthers a resource-based account of knowledge, reasoning, and learning.

We begin with the construct of *epistemological framing*, which we develop from previous work on framing and student epistemologies. We then present the behavioral clusters evident in tutorial video data and our arguments for their connection to epistemological framing.

## II. THEORETICAL FRAMEWORK

### A. *Epistemological framing*

*Framing* is a construct developed in anthropology and linguistics[1] to describe how an individual or group forms a sense of "What is it that's going on here?" (Bateson, 1972; Goffman, 1986; MacLachlan & Reid, 1994; Tannen, 1993). To frame an event, utterance, or situation in a particular way is to interpret it based on previous experience: to bring to bear a structure of expectations about a situation regarding what could happen, what portions of the information available to the senses require attention, and what might be appropriate action. For example, monkeys engaged in biting each other are skilled at quickly and tacitly "deciding" whether the biting is aggression or play (Bateson, 1972). An employee may frame a gift from her supervisor as kind attention or as unwelcome charity. A student may frame a physics problem as an opportunity for sensemaking or as an occasion for rote use of formulas.

In school settings, epistemological framing is of particular importance: students form a sense of what is taking place with respect to knowledge, including, for example, what portions of information and experience are relevant for completing assignments. Other aspects of framing are important as well, including social framing, in which students form a sense of what to expect of each other, of their instructor, and of themselves. For groups of students working together collaboratively, different aspects of framing interact.

Previous analyses have found evidence of framing primarily in linguistic markers associated with speech acts.[2] Nonetheless, because frames "emerge in and are constituted by verbal and nonverbal interaction" (Tannen, 1993, p. 60), evidence of framing can come in a variety of forms. In our work we have found a wealth of paraverbal and nonverbal cues of framing, and so have focused on those cues in addition to linguistic markers. In any social interaction, framing presents a communicative task in which participants collaboratively establish the nature of their shared activity (Tannen, 1993, p. 66). Conversational partners both indicate the nature of the activity they're engaged in by means of verbal and nonverbal interactions, and observe others' behavior in order to learn what kind of activity is taking place. If a speaker makes eye contact, gestures while speaking, and uses an animated voice, she is both experiencing an engaged discussion herself and displaying to others that they are mutually engaged in a discussion.[3] Participants interpret one another's signals regarding the nature of the activity by means of contextualizing cues that include body language (*e.g.,* facial expression, gesture, and posture), prosodic features of utterances (*e.g.,* pitch variation, loudness, pausing, pacing), and linguistic signals (choice of vocabulary, level of formality, choice of pronouns) (Goodwin, 2000; Tannen, 1993, p. 62).

For example, in an excerpt from students working on a tutorial we quote and analyze more extensively below, Jasmin[4] says:

> *In summary, for most people* blah blah blah blah... *We shouldn't dwell on this kind of question and instead focus on learning exactly when...* This is like one of the questions.

On video, we observe a close correspondence between the substance of Jasmin's speech and her physical behavior. The italicized portion of her statement consists of abbreviated but otherwise verbatim statements from the worksheet. While she says those words, Jasmin is leaning over and looking at the worksheet, using a singsong tone, speaking softly, using non-words ("blah blah"), and hugging her arms closely to her body. Her behavior matches the evidence from her speech, which is that she frames the activity as form of *reading aloud from a common text.* Jasmin's expectations for this activity are indicated both physically and verbally: When you read, you look at what you are reading. A singsong tone and lack of gestures signal recitation (Gerwing & Bavelas, 2004; Scherr, in press). If one's peers are reading the same text, it is appropriate to speak quietly and abbreviate, since the point is more to indicate what you're reading than to insure others' comprehension.

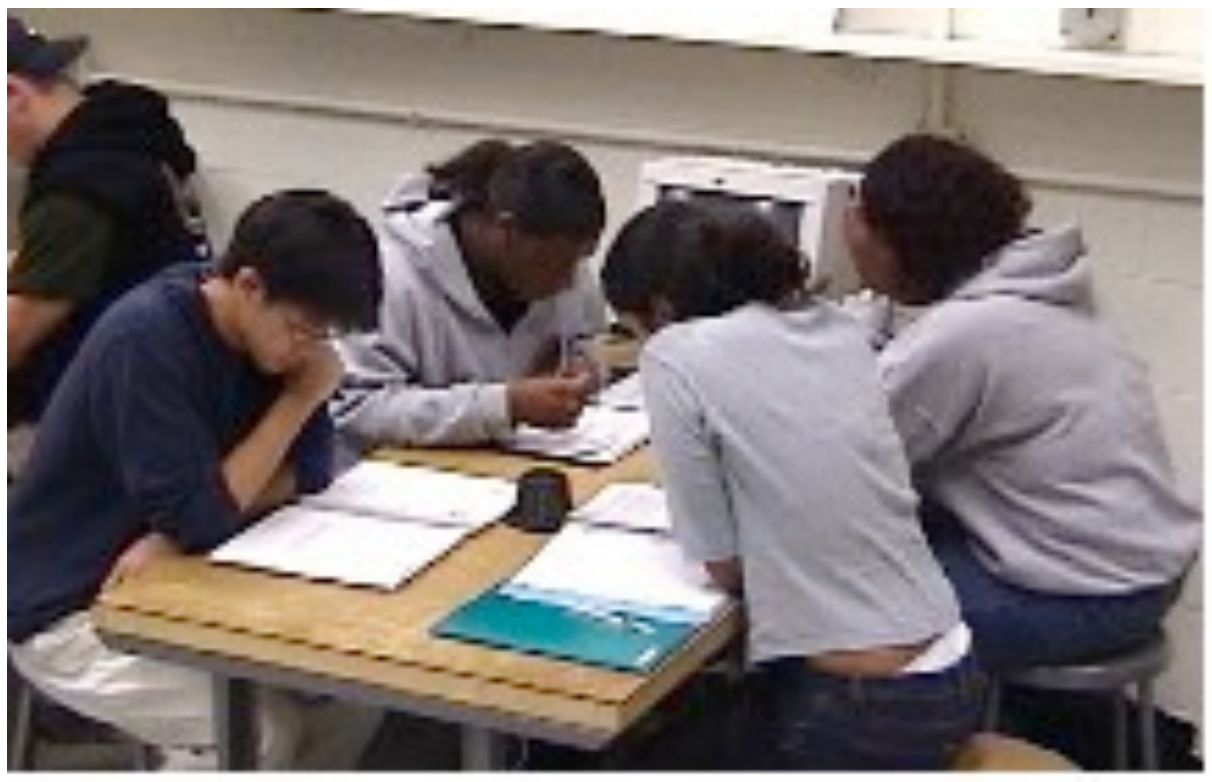

Figure 1. Nonverbal cues indicate reading from a shared text.

Then, when Jasmin says, "This is like one of the questions," she speaks more loudly without the singsong recital tone, pushes back from the table, and points at her paper with her pencil. These paraverbal and nonverbal behaviors help her peers to recognize that she's doing something different now: Jasmin frames the new activity as *commenting on what she's just read* (in particular, pointing out a similarity between the worksheet task and another question she expects is familiar to her peers). Again, her behavior matches the substance of what she is saying. Her clear, loud voice denotes original speech rather than recitation; she indicates that she is now talking *about* the worksheet (rather than reading from it) both by pointing to it and by distancing her body from it. One of her partners, Sheryl, recognizes this change in the moment and knows what she's referring to: she sits up, looks at Jasmin, and says, "Yeah, this is question six on the homework."

Humans are continually confronted with interpretive tasks such as the one Sheryl faced in understanding Jasmin. In order to understand what an utterance means, we have to sense what kind of activity the utterance is part of: that is, we have to frame the activity. Is it reading, commenting, explaining, imitating, chatting? We make this interpretation based on verbal, paraverbal, and nonverbal cues in the interaction – based not only on *what people say* but also on *how they say it* – and we think and act in response to the interpretation we make.

In the examples of behavioral analysis that follow, we articulate a dynamic between readily observable behavior and epistemological framing. In doing so, we make a methodological case that observations of physical behavior can provide evidence of epistemological framing, as well illustrating how epistemological framing can promote particular kinds of behavior.

### *B. Continuity between cognitive and group dynamics*

The other task of this paper is to articulate the dynamic between epistemological framing and the substance of student thinking. This interplay between students' behavior, conceptual reasoning, and epistemological framing calls for theoretical continuity between individual cognition and group dynamics.

A *resource model* of knowledge and reasoning provides this continuity. Our larger program of research to develop resource models (Hammer, 2000; Redish, 2004; Scherr, 2007) is in line with diSessa's (1993) model of intuitive physics as made up of fine-grained, context sensitive "phenomenological primitives" and, more generally, with Minsky's (1986) model of mind as consisting of manifold cognitive "agents." Hammer and Elby (2002) proposed modeling intuitive epistemologies in terms of epistemological resources, rather than using beliefs, theories or stages of development as the units of cognitive structure. That is, rather than attributing unitary epistemologies to students, such as that knowledge is received from authority, certain, or simple (see review in Hofer & Pintrich (Eds.), 2002), this view attributes a variety of resources for thinking about knowledge in different ways at different times. From early ages, students have resources for thinking of knowledge as received, as inferred from other knowledge, as invented or as perceived. They have resources as well for understanding a wide variety of kinds of knowledge, epistemic activities, and stances—facts, stories, lists, rules, suppositions, doubt, certainty, and so on—resources they may activate or not depending on the particular situation. Thus a resource-based view of epistemologies accounts for the variability and multiple coherences in student reasoning, as evident in studies focused on individual students across different contexts (Knight & Sweeney, under review; Leach, Millar, Ryder, & Sere, 2000; Lising & Elby, 2005) and on groups or classes (Louca, Elby, Hammer, & Kagey, 2004; Rosenberg, Hammer, & Phelan, 2006; Roth & Roychoudhury, 1994; Sandoval, 2003).

Redish (2004) proposed the idea of *epistemological frames* to connect this work to the construct of framing in linguistics and anthropology. At the level of resources, we take a frame to be a locally coherent pattern of activations – "coherent" in that the pattern holds together for some length of time and "local" in that the coherence may be particular to the moment or context (Hammer, Elby, Scherr & Redish, 2005). This is a dynamic systems account of framing; coherences emerge from the activations and interactions of many cognitive elements. They may involve resources within an individual's mind or across multiple individuals or a group. Framing is defined in the literature in terms of individual reasoning (Tannen, 1993) and of social dynamics (Goffman, 1986); a resource-based model of framing smoothly links this discourse dynamics with models of individual cognition (Brown & Hammer, in press). In the examples of behavioral analysis that follow, we show connections between the substance of individual students' thinking and the nature of interactions among members of the group.

## III. CONTEXT FOR RESEARCH

The tutorials took place as part of a two-semester algebra-based introductory physics course at the University of Maryland, with approximately 160 students in each lecture section, most of whom are junior and senior health and life science majors. More than half are female and there is wide ethnic diversity reflecting the student population of the University of Maryland. The course was reformed as part of a project titled *Learning How to Learn Science: Physics for Bioscience Majors,* carried out at the University of Maryland from 2000-2005. The goal of this project was to determine whether an introductory physics course could serve as a venue to help biology students learn to see science as a coherent process and way of thinking, rather than as a collection of independent facts, and whether this goal could be achieved within the context of a traditional large-lecture class without a substantial increase in instructional resources. The project adopted reforms that were well-documented to produce conceptual gains and adapted them to try to create a coherent package that produced epistemological and metacognitive gains. We hoped that this could be done without sacrificing the conceptual gains associated with these reforms and this indeed turned out to be the case.[5]

As part of the course reform, the traditional teaching-assistant-led recitation was replaced with worksheet-based group-learning activities ("tutorials") based on the model developed at the University of Washington (McDermott et al., 1998; McDermott, Shaffer, & Somers, 1994). In the tutorial sessions, students worked in small groups on worksheets that led them to make predictions and compare various lines of reasoning in order to build an understanding of basic concepts. Teaching assistants (TAs) served as facilitators rather than as lecturers: instead of demonstrating problem solutions to students, they engaged the students in discussions of difficult concepts. Each class section consisted of five groups of four students each, supervised by two TAs. The tutorials were constructed to emphasize the reconciliation of everyday, intuitive thinking and experience with formal scientific thinking and to encourage explicit epistemological discussions about the learning process (Elby, 2001). They are published as part of a project whose purpose is to provide instructors with the materials and resources to implement similar tutorials at their own institutions (titled *Helping Students Learn How To Learn: Open-Source Physics Worksheets Integrated With TA Development Resources*).

The *Helping Students Learn How To Learn* project made extensive use of video of students working on these tutorials, both to assess the effectiveness of the tutorials and to develop resources for instructors who wanted to implement them. Approximately 380 hours of student tutorial groups were collected during the *Learning How To Learn Science* project. From that library, the *Helping Students Learn How To Learn* project selected videotapes from a single academic year in which the students rated the tutorials highly, most of the TAs were members of the Physics Education Research (PER) Group experienced with tutorial instruction, and the lecturers were PER experts who supported the tutorial part of the curriculum. From the collection of tapes from that academic year, the project selected seven groups that observers identified as being "watchable": They were consistently on-task and talked to each other frequently.[6]

For the study reported in this paper, we selected three of the best-tested tutorials from the set of tutorials published in the *Helping Students Learn How To Learn* project – tutorials that cover basic topics common to most introductory physics curricula (Newton's third law, free-body diagrams, and electrostatics). We randomly selected three student groups from the groups published with that project. We represent the student groups with letters (A, B, and C). We analyzed group A doing the Newton's third law tutorial and a tutorial on free-body diagrams, group B doing the tutorial on electrostatics, and group C (in the same classroom and with the same TA as group A) doing the tutorial on Newton's third law. The final data set for this study thus consists of four hours of tutorial videotape documenting three different groups of students working on three different tutorials. In what follows we describe in some detail our analysis of group A working on the Newton's third law tutorial, and give one example from our analysis of group B working on the electrostatics tutorial.

In selecting the data for the study presented in this paper, we did not attempt to create a representative set of videotapes. Instead, our data set represents a small set of examples drawn from a "best-case scenario": a carefully designed, well-functioning class with experienced TAs, supportive instructors, well-tested tutorials, and relatively engaged students.

## IV. IDENTIFYING AND INTERPRETING BEHAVIORAL CLUSTERS

As mentioned in the introduction, we have found four distinct patterns of students' physical behavior in the data for this study. In what follows we present the method we used to identify these behavioral clusters and the content of each of the four clusters that we identified.

### *A. Methodology*

In order to identify clusters of behaviors that appear in our data set, coders independently watch video of small groups of students and note the exact time of *changes in* the students' vocal register, affect, grammar, gesture production, and body language. Having noted changes in behavioral clusters, we then systematically identify the features of the behavioral cluster: hand motions including gestures (Goldin-Meadow, 2003), facial aspect, body position and/or movement, vocal register, gaze, and so on. We label the behavioral clusters with meaning-neutral labels (colors). Identifying the behaviors within each cluster permits coders to reliably observe when students participate again in a behavioral cluster the coder had previously identified (*i.e.,* shift back into a certain cluster after engaging in other behaviors).

Coding is performed in real time without transcript. At least two researchers code the changes independently. Tannen's (1993) research led us to expect that clusters of behaviors typically change all at once for an entire group of participants, and that is generally the case in our data. Coders are able to code the whole group reliably (peers typically share the same behavioral cluster[7]). The behavioral clusters are distinct and easy to identify, with 90% inter-rater reliability before discussion, to five seconds' accuracy.

Working in this way, we identified four clusters of behavior that applied to all three groups we studied. The fact that the behavioral clusters identified for one group appeared in others as well is a finding in itself, probably reflecting a set of cultural norms shared by the participants (almost all of whom are junior and senior university students).

### *B. Behavioral clusters defined*

1. Blue behavioral cluster

In the behavioral cluster that we have labeled "blue," students' eyes are primarily on their papers with brief glances up at their peers. Their bodies lean forward at an angle of about 30° to the vertical, their hands are mainly at rest (few gestures), and their faces are relatively neutral; Figure 1 is typical. Their tone of voice is low, quiet, and indistinct, and their speech consists largely of phrases read from their worksheets; when they speak to one another, they briefly and quietly speak with rising intonation, and answer briefly and quietly with falling intonation. The initiator often utters an incomplete sentence that is completed by a peer.

2. Green behavioral cluster

In the "green" behavioral cluster, students sit straight up and make frequent eye contact with one another, as shown in Figure 2. Their faces and voices are animated; they gesture relatively prolifically and engage in clear, loud, original speech.

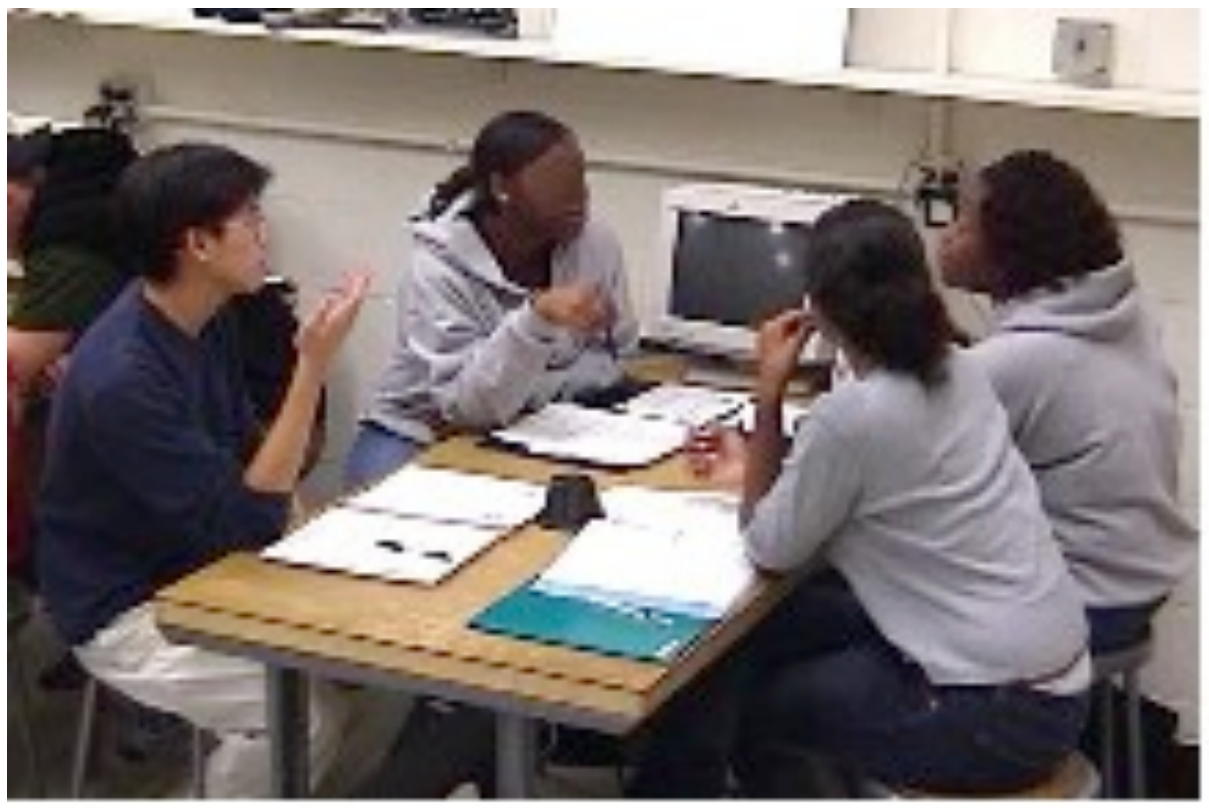

Figure 2. Students in the green behavioral cluster typically sit straight up, make eye contact, and gesture.

3. Red behavioral cluster

The "red" behavioral cluster involves the students' interaction with a teaching assistant, although students need not exhibit the red behavioral cluster all the time that a TA is present.[8] The students speak little and make eye contact with the TA. They sit straight with their bodies still, and their hands are relatively quiet except for habitual movements.

4. Yellow behavioral cluster

In the "yellow" behavioral cluster, students giggle or smile, and have hedging or joking tones of voice. They shift their bodies around in their seats and touch their own faces and hair. Their gaze is unsettled, moving among peers, papers, and other points in the room. The yellow behavioral cluster appears to be somewhat less stable than the other three clusters identified in our examples: although it recurs regularly, it rarely lasts even as much as thirty seconds, whereas the other three clusters often persist for minutes at a time.

A summary of the features of the four behavioral clusters appears in Table I.

TABLE I. Features of the four observed behavioral clusters.

| **BLUE** | **GREEN** |
|---|---|
| Eyes on paper or glance at peers<br>Bodies lean forward about 30°<br>Hands at rest or holding pencil<br>Neutral tone and face<br>Voice low, quiet, indistinct<br>Read aloud or briefly consult peers | Eye contact with peers<br>Sit up straight<br>Prolific gestures<br>Animated tone and face<br>Voice clear, loud<br>Original speech |
| **RED** | **YELLOW** |
| Eye contact with teaching assistant<br>Sit up straight<br>Hands quiet or reduced gestures<br>Minimal speech | Unsettled gaze<br>Fidget in seat<br>Self-touch, shrugging<br>Hedging or joking tone; giggling, smiling |

*C. Behavioral clusters interpreted*

We labeled the behavioral clusters with colors, in an attempt to suspend our inclinations to interpret them while we defined the specific physical behaviors involved. Having arrived at these definitions and shown inter-rater reliability, we shifted to interpretation; what do these different behavioral clusters suggest with respect to epistemological framing? For us, as we expect for the students, this interpretation is automatic and unconscious—it is difficult not to do it. That is reason for attending to it systematically in analyzing student thinking.

The behaviors associated with the blue and green behavioral clusters indicate very different framings of the tutorial activity, in spite of the fact that both might loosely be called "working collaboratively on the tutorial." In the blue behavioral cluster, the fact that students' eyes are on their papers and their bodies lean over the desks sends the message that their main interaction is with the worksheet. Students' neutral facial expressions and lack of gestures further suggest that they expect minimal interaction with their peers; the tutorial activity is, apparently, primarily expected to be individual at this point. They do glance occasionally at one another, indicating an expectation of "check-ins" with peers from time to time, and they read aloud, indicating that they expect their partners to coordinate with them. The reading aloud, however, is indistinct, indicating that they expect their peers are already reading the same thing on their own. The students evidently frame the activity as primarily *completing the tutorial worksheet.*

The green behavioral cluster, on the other hand, is characterized by original speech delivered in relatively loud, animated voices, indicating intellectual and/or emotional engagement. Sitting up straight and making eye contact with one another, students display an expectation that their attention belongs on their peers. Their clear speech and prolific gestures indicate further that they expect their peers to pay attention not only to the fact that they're talking, but also to the details of what they're trying to express. Watching students behave in this way, we understand them and they understand each other to be engaged in a *discussion.*

In the red cluster, students display expectations that their attention belongs on the TA, suggesting that they frame the activity as *listening to the TA.* Variations of the red behavioral cluster sometimes include students responding to TA requests for explanations of phenomena that the students had already discussed among themselves. In those cases, the red behavioral cluster includes students making more original contributions to the conversation, as well as gesturing. The verbalizations and gestures, however, have a different quality than those observed in the green frame; the gestures, for example, are often reduced versions of gestures originally made during the green behavioral cluster (Scherr, in press). Students' attention remains primarily on the TA, as indicated by gaze and body orientation. In those cases, however, it might be more accurate to label the activity as *responding to* (rather than *listening to*) *the TA.*

The behaviors associated with the yellow behavioral cluster are less straightforward to interpret than those of the other clusters. On the one hand, they seem to indicate social discomfort or perceived vulnerability: students avoid direct eye contact, fidget in their seats, and fiddle with objects. On the other hand, they laugh, smile, and appear to joke with one another, behavior that indicates play. These characterizations are not mutually exclusive; play is a "nonliteral orientation" (Tannen, 1993, p. 114) that may be initiated as a cover for actions involving some personal risk. This behavioral ambiguity corresponds to ambiguity in the communicated substance of student thinking; we describe an example below. We refer to the frame indicated by the yellow behavioral cluster as the *joking frame* to include the sense that students are signaling that the content of their speech should not be taken literally (even though they may also intend, on some level, to explore the ideas that they express; see below).

Table II details the behaviors characterizing the different behavioral clusters and the expectations we take as indicated by those behaviors. To be clear, we are describing our interpretations as what we expect is shared cultural knowledge, shared among students, researchers, and most readers of this journal. The analyses of specific episodes below will support these interpretations.

## V. ANALYSES OF STUDENT BEHAVIOR AND EPISTEMOLOGICAL FRAMING

In what follows we describe in detail some of the shifts in behavioral clusters that we observed in our analysis of the selected groups and tutorials, and we show that those shifts correspond to shifts in the substance of students' reasoning as evident in their speech and gestures. Taken together, the students' behavior, speech and gestures support interpretations of the students' epistemological framing.

### *A. First episode: Group A on Newton's third law*

In the first three minutes of working actively on the tutorial worksheet on Newton's third law, the students in group A consider the first several questions that the worksheet poses. The tutorial begins by stating Newton's third law and admitting that in some cases it seems not to make sense (an admission that is well supported by research into student understanding of Newtonian mechanics (Boyle & Maloney, 1991; Hestenes, Wells, & Swackhamer, 1992; Maloney, 1984)). The worksheet asks students to consider a heavy truck ramming into a parked, unoccupied car: “According to common sense, which force (if either) is larger during the collision: the force exerted by the truck on the car, or the force exerted by the car on the truck?” The instructional activity is described in detail in (Elby, 2001). Figure 3 shows the tutorial excerpt containing the questions they’re discussing.

TABLE II. Specific behaviors found in each cluster and expectations likely to be associated with those behaviors.

| **BLUE:** *Worksheet frame* | | **GREEN:** *Discussion frame* | |
|---|---|---|---|
| Behavior | Expectation | Behavior | Expectation |
| Hands quiet, face neutral | Minimal interaction, individual activity | Prolific gesturing | Peers are watching and want to understand |
| Body leans forward, eyes on paper | Attention belongs on the worksheet | Animated tone, face | Intellectual and/or emotional engagement |
| Brief glances at peers | “Check-ins” expected | Sit up straight, eye contact | Attention belongs on peers |
| Read aloud without clarity | Peers coordinate, but read on their own | Clear, original speech | Peer interest in details of thinking |
| **RED:** *TA frame* | | **YELLOW:** *Joking frame* | |
| Behavior | Expectation | Behavior | Expectations |
| Sit up straight; eye contact with TA | Attention belongs on TA | Giggle, smile, self-touch, fidget, unsettled gaze | Embarrassment, perceived vulnerability |
| Reduced gestures | Rehashing thinking | | |

In the episode below, “Kendra” (K) expresses a common concern about whether the third law applies to the situation described; “Alan” (A) and “Jasmin” (J) try to reconcile the third law with the common-sense intuition that the force by the truck would be larger. “Sheryl” (S) is present but does not speak in this episode. Italics denote quotations from the tutorial worksheet.

A: *According to common...*
K: Yeah, I think I don’t understand this.
A: *According to common sense, which force, if either is larger during the collision: the force exerted by the truck on the car, the force exerted by the car on the truck?* Oh...
K: I mean, the truck.
A: The truck, yeah.
A: OK, so...
J: *According to common sense, which force is larger in the collision?* Tuh...
A: What?
J: Why are they asking that?
A: Oh, OK, the car...
J: question...
A: *Is your group’s explanation in Part A similar or different?* It’s similar.
J: *Intuitively, the car reacts more, you’d rather be riding in the truck, so the car feels... is your group’s explanation similar or different?*
A: It’s the same. Right. OK, so, *according to Newton’s Third Law, which of the... which of those forces...*
K: It says they’re equal, right?
A: Yeah. Should be equal.
K: I kinda, I could never understand that, but.

K (cont.): Does this go against the law then, or is it that they are equal but we just think it's the truck? You understand what my question...

A: We think it's the truck because the truck doesn't move backwards, I think. Right? Cause if there's equal and opposite forces, the truck... we would, if we actually saw it, we'd think the truck would hit the car and go backwards because of the force, but since...

J: Maybe they do exert the same force, but the truck doesn't move.

A: The truck doesn't move cause I think it's got the momentum going, and... you know.

K: So they, they are doing the same force, it just doesn't... it's just not common sense.

A: Yeah, it just doesn't register because we see the truck (K: It looks as being bigger) still moving forward, right, right.

> The main point of this tutorial is helping you learn more strategies for learning physics concepts that seem to defy common sense.
>
> **I. Newton's third law and common sense**
>
> According to Newton's third law, when two objects interact,
>
> *The force exerted by object A on object B is equal in strength (but opposite in direction) to the force exerted by object B on object A.*
>
> Often, this law makes perfect sense. But in some cases, it seems not to.
>
> Consider a heavy truck ramming into a parked, unoccupied car.
>
> A. According to common sense, which force (if either) is larger during the collision: the force exerted by the truck on the car, or the force exerted by the car on the truck? Explain the intuitive reasoning.
>
> B. We've asked this question of many students, and a typical response goes like this:
>
> "Intuitively, the car reacts more during the collision. (You'd rather be riding in the truck!) So the car feels the bigger force."
>
> Is your group's explanation in part A similar to or different from this? Explain.
>
> C. According to Newton's third law, which of those forces (if either) is bigger?
>
> D. Experiment. Is this a case where Newton's third law doesn't apply? At the front of the room, the TA has set up an experiment that simulates a truck ramming a car. Go do the experiment and record the results here. You can also test whether Newton's third law holds for other collisions.

Figure 3. First excerpt of tutorial on Newton's third law.

1. Behavioral coding

In the first part of the episode (lines 1-29), students lean over their papers and only briefly glance up at their peers. They rarely gesture, and their faces and voices are relatively neutral. Their speech consists largely of phrases read from their worksheets. Overall, their behavior is well-characterized by the blue behavioral cluster. At the moment indicated by the horizontal line in the transcript (beginning at line 30), however, the students' behavior changes abruptly to that of the green behavioral cluster: They sit up and make eye contact with one another, make original statements in clear, loud voices, and gesture relatively prolifically. The green behavioral cluster continues to the end of the episode.

There are exceptions to the above characterization of student behavior in the episode. Right at the beginning of the episode (line 2), Kendra's statement, "Yeah, I think I don't understand this," is unusually clear, original speech compared to the other utterances that take place shortly before and after it; she is not reading from the worksheet, and her voice is relatively loud and animated. Other aspects of her behavior, though, remain consistent with the blue behavioral cluster – for example, she remains bent over her worksheet. Kendra's statement appearing immediately before the shift from blue to green, "I could never understand that" (line 28), is similar in both content and tone to the slightly anomalous statement at the start of the episode. We interpret the momentary contrast between her behavior and the group's in the next section.

2. Substance of student thinking

The contrast between the blue and green behavioral clusters is reflected in the substance of students' speech. While students are exhibiting the "blue" behaviors, their speech consists mostly of reading worksheet questions aloud. They offer only brief answers to or comments about the questions, and they ask only two questions of their own: one to verify an answer ("It says they're equal, right?") and the other focused on the worksheet's intention ("Why are they asking that?").

The green mode begins with Kendra's question about the physics ideas: does this situation violate Newton's third law, or is there some misunderstanding? Alan responds with a perceptual reason why we might think the truck's force was larger (it "doesn't move backwards"), implicitly asserting that the forces are equal; his explanation is incomplete, perhaps indicating that he's constructing it as he speaks. Jasmin adds to Alan's reasoning by suggesting that perhaps Newton's law holds in this situation ("they do exert the same force") but it's hard to tell because "the truck doesn't move" (by which she might mean that the truck's motion is relatively unaffected). Alan gesturally "waves off" the collision's lack of effect on the truck with reference to what he seems to identify as a different phenomenon ("momentum"). Kendra tries to affirm that the vehicles exert equal forces on one another in spite of common-sense convictions to the contrary. Alan agrees, and asserts that one may be misled by perceiving little change in the motion of the truck ("it just doesn't register because we see the truck still moving forward").

3. Epistemological framing

The changes in behavioral cluster from blue to green correspond with changes in the substance of the conversation. These two types of evidence together support our identification of the students' epistemological framing: They shift from framing their activity as *completing the worksheet* to framing it as *discussing the ideas*. This is a local shift; we expect these two frames are part of a larger framing of what is taking place in this tutorial session as a whole. Studying student behavior at this grain-size, we believe, provides insight into that larger framing as well as into its dynamics, including the coordination of the *worksheet* and *discussion* frames.

This identification of a shift in epistemological framing provides a basis for understanding Kendra's locally anomalous behavior in lines 2 and 28. Kendra's initial statement may have been a *bid* for the group to change its activity; the group accepted her bid the second time she made it. We identify her bid for a frame change based entirely on her behaviors (*e.g.,* she is briefly more animated). The content of her speech, however, is also similar in the two cases: she denies understanding the situation. Her peers are likely to interpret such a statement as an implicit request for understanding. Such a request is more consistent with framing the activity as a discussion than with framing it as filling out a worksheet.

*B. Second episode: Group A on Newton's third law, continued*

Shortly after the episode described above, the same students consider the next question in the tutorial. This question is epistemological: it prompts the students to reflect on their own views of knowledge and learning. The question is reproduced in Figure 4.

**II. What to do with the contradiction between common sense and Newton's third law?**

Before moving on to the next part of our Newton's third law lesson, let's consider the contradiction we just found between physics and common sense.

A. In summary, for most people, Newton's third law contradicts the common-sense intuition that the car reacts more during the collision. Which one of the following best expresses your attitude toward this contradiction?

 i. We shouldn't dwell on these kinds of contradictions and should instead focus on learning exactly when Newton's third law does and doesn't apply.

 ii. There's probably some way to reconcile common sense with Newton's third law, though I don't see how.

 iii. Although physics usually can be reconciled with common sense, here the contradiction between physics and common sense is so blatant that we have to accept it.

Briefly explain why you chose the answer you chose.

B. Discuss your answer with your group. Is there a consensus or do people disagree?

Figure 4. Second excerpt of tutorial on Newton's third law.

In the episode below, Kendra asserts that in this case the contradiction between physics and common sense is so extreme that there's no way to reconcile the two. Jasmin and Alan have more faith that reconciliation is possible, and try to help Kendra see how the contradiction might be resolved. A TA, who has been eavesdropping on the conversation, arrives and compliments Kendra on her willingness to state that she doesn't believe Newton's third law can apply to this situation. Again, statements that students read verbatim from the worksheet are italicized.

J: *In summary, for most people* blah blah blah blah... *We shouldn't dwell on this kind of question and instead focus on learning exactly when...* This is like one of the questions.

S: Yeah, this was question 6 on the homework.

A: Really?

J: Mmmhmm.

A: I haven't gotten that far.

(pause)

K: Why? Six is just an explanation question?

S: Mmmhmm.

K: Oh really?

S: That's why I said that was good.

A: What'd you put? Two? Three?

S: Huh?

A: Three?

S: What do you mean?
A: You put three? A?
S: Oh yeah.
A: You put three too?
K: Hell yeah.
J: I put two.
K: I'm mad, so I would have picked three.
S: Oh, you picked two?
A: Yeah, because, I mean...
J: That's what he even said in class, there's always some way to like...
A: There's always some way to explain

K: You think so?
J and A: Yeah.
K: I don't think there's any way that you can explain to me how a massive truck is going to have the same forces on it...
J: I mean, I think they do feel the same force... (K: I just have to accept it), it's just one is like...
K: I don't see it!
J: Just... it just looks like that one got the push.
A: Yeah, it always looks like the one gets pushed...
K: No, like...
A: I'm sure the truck gets like...
K: Think of a Porsche versus a humongous truck, (S: I know, that doesn't make sense) how do you explain to me...?
A: OK, the Porsche would get totaled, right? But then I'm sure the truck's...
J: Felt the same.
A: The fender probably got like, you know, messed up.
K: Scratched?
A: But you know, it still did get some damage, so, I mean, there's still some...
K: No.
A: No?
K: No.
A: It's not working?
K: I just have to accept it.
TA: So, this is cool. I just heard that you're having trouble getting, like, you're saying no, I can't believe that.
K: Yes.
TA: I can accept it, but I can't believe it. And it's cool that you can realize that, you know, if that's how it is, but I don't get that. That's what we're gonna work on later through this tutorial. We're gonna try to understand how...
J: Yeah, why that happened.
TA: Yeah.
K: OK.
TA: But it's cool that you could realize that "I don't get that."
J: It's cool.

1. Student behavior

During the first minute of this episode (lines 1-32), students exhibit the blue cluster of behaviors: eyes primarily on their papers, bodies leaning forward, hands mainly at rest, faces relatively neutral, and vocal tone low, quiet, and indistinct. There are two brief exceptions, first when Jasmin says that "this is like one of the questions" (lines 4-5); she speaks more loudly, glances up at her peers, pushes back from the table, and points at her paper with her pencil. The second occurs when Kendra says, "Hell yeah" (line 24), a statement that is well outside the norm of indistinct speech that characterizes the blue behavioral cluster.

The group's behaviors do not change, however, until a few lines later, when Kendra says, "You think so?" (line 32). Kendra's two statements (lines 24 and 32) may again be characterized as bids for frame change, only the second of which is successful. After line 32, the students participate in behaviors characterizing the green behavioral cluster: they sit up, make eye contact with one another, gesture relatively prolifically, and have animated faces and voices.

The TA's arrival at the table precipitates another change, now to behaviors identified with the red behavioral cluster. The students stop talking; their eye contact is with the TA; they sit straight with their bodies still; and their hands are relatively quiet except for habitual movements (Jasmin plays with her necklace, and Alan fiddles with his pen and paper). For part of the time that the TA is present, Sheryl leans forward and writes on her worksheet, behaviors that are typical of the blue behavioral cluster rather than the red. However, we recognize her as behaving outside the designated activity of the moment. In some ways, her contrasting behavior draws our attention to the dominant group activities, which are those of the red behavioral cluster (Goffman, 1986, Ch. 7).

2. Substance of student thinking

Again, we observe that the substance of students' discussion changes markedly along with their behaviors. While the students engage in the behaviors associated with the blue cluster, the main substance of the students' speech (other than conferring about the homework) is consulting one another about their answers to the multiple-choice question ("What'd you put? Two? Three?"); Jasmin refers to "what he even said in class" as justification for answer 2.

As students shift to the green cluster, Kendra challenges the idea that the collision forces are in fact equal; she is willing to accept the result as fact, but asserts that it is beyond reason ("I don't think there's any way that you can explain [it] to me"). Alan offers an explanation in terms of damage to the truck ("the fender probably got messed up"), an explanation which appears to strike Kendra as weak based on the incredulous tone of her reply ("Scratched?"). Alan acknowledges that he's not persuading her ("It's not working?") and Kendra states that she's "just going to have to accept it." The dialogue is original, and its substance is both intellectually demanding and emotionally charged; it is easy to distinguish from the relatively rote exchanges that take place in the worksheet frame.

When the students shift to the red cluster of behaviors, they say little (as is generally the case for that cluster), and we have little useful evidence of their reasoning other than their stated agreement with the TA's statements.

3. Epistemological framing

Again, we see a correspondence between the students' behavior and the substance of their speech. The blue cluster, with its physical indications of a focus on the worksheet, corresponds to brief comments and questions about the answers and, again, a reference to authority. The green cluster, with its physical indications of students' attending to each other, corresponds to discussion about ideas. Here the transition occurs at line 32, a shift to what "you think" from what the professor "said in lecture" (29-30). Kendra challenges the idea that the force could be equal, and the group takes that up as a topic. In this way, the students' behavior and words support the claim that they shifted from framing their activity as *completing the worksheet* to framing it as *discussing ideas.*

At a larger grain-size, we might understand this group as framing tutorials in general as a collection of activities for learning the concepts of the course, of making progress through the worksheet while monitoring their understanding for gaps or discrepancies that require attention. That interpretation would be supported by evidence to establish that the patterns in the two episodes from this group applied more generally across their behavior and conversation in multiple tutorials.

When the TA arrives, there is another shift. The red cluster, with its physical indications of the students' receptiveness to the teaching assistant, corresponds to the students providing only affirmation of his remarks. Their bodies and faces are turned attentively to the TA, and they are mostly quiet; their minimal speech appears to be entirely aimed at acknowledging the TA's speech ("Yes," "Yeah, why that happened," and "Okay"). Both behavior and speech indicate that they are framing the activity as *listening to the TA.*

*C. Third episode: Group B on electrostatics*

We turn now to a different group working on a different tutorial, in order to provide an example of the yellow behavioral cluster, which we did not observe in the above episodes with group A. Group B was working on the tutorial on electrostatics. Early in the tutorial, students rubbed a Styrofoam picnic plate with a piece of wool cloth and observed that the plate became charged as a result. In the following episode, students discuss an exercise requesting them to draw a diagram showing the charge on the plate and on the cloth after rubbing. The exercise is shown in Figure 5.

> Suppose you were to take one of the foam plates and rub it with a piece of cloth. Draw a diagram showing the charge on the plate and on the cloth. Use "+" and "–" symbols to indicate the parts that are charged and the type of the charge. (Can you tell whether the plate is + or –? If not, just pick one!)

Figure 5. Excerpt of tutorial on electrostatics.

In this episode, student discussion focuses on which object is positive and which is negative. "Megan" (M) proposes that it's not possible to tell which is which, and "Vicki" (V) initially agrees with her. "Dalia" (D) thinks a test probably exists. "Chris" (C) offers a guess that the plate is positive; when Vicki says she would have guessed the opposite, Chris gives a half-joking explanation of why the cloth would acquire negative charge more readily than the plate. The rest of the group is amused by his idea, and Vicki offers to adopt his answer.

M: *Take one of the foam plates and rub it with*

V: *Take one of the foam plates and rub it with a piece of cloth.*

D: Here's the plate; here's the cloth. [Drawing.]

V: How could we tell if it's positive?

M: *Can you tell whether the plate is* – I don't think you can tell.

V: I don't think you can tell.

C: Is there a rule that dictates which way the electrons are going to go?

D: There's probably a way to test it, like somehow but –

V: I'm sure it depends on the material.

M: There's no way, there's no way to tell.

V: But I don't think we know right now.

M: Yeah.

D: Right. That's what I'm sayin'. All right that's plus that's minus. [Drawing.]

C: I mean, I'd say the plate, just guessing, I'd say the plate's positive but I don't have a reason for that. I'd say that –

---

V: I was going to say the plate was negative, but that was just guessing. [Laughter.]

C: I like, I would say that, it just seems to me that electrons. This [the cloth] looks like it could hold electrons to me. [Laughter.]

V: ‘Cause its soft. There’s room for them. [Laughter.]
C: It’s soft. [Rubbing the cloth on his face.] Yeah.
[Laughter from group,]
D: They’d slide off the plate.
V: It’s roomier.
C: It’s quilted, you see.
D: It’s like toilet paper. [Laughter from group.]
V: All right we’ll make the plate positive. All right?
C: Thank you. Appreciate that. [Laughter.]

1. Student behavior

In the first part of the episode (lines 1-24), students exhibit the behaviors characterizing the blue behavioral cluster. Just as with group A, the students lean forward, eyes mainly on their papers, speaking in soft voices with little body movement or gesturing. In this episode, however, the students do not exhibit behavior in the green cluster. Rather, after line 26 we see the set of behaviors identified with the yellow behavioral cluster: the students giggle and smile, use noncommittal or gently mocking tones of voice, move around in their seats, and touch their own hair or face. Their gaze shifts among peers, papers, and other points in the room.

2. Substance of student thinking

While students exhibit the behaviors in the blue cluster, the students have more of a conversation than we saw in the episodes above of group A; they do not read as much from the worksheet. The substance of the conversation, however, is limited to the instructions for what to do (1-5), and questions and comments about the possibility of answering the worksheet questions. They ask each other “how can we tell” and “is there a rule that dictates”; they comment about whether there might be “no way to tell” or venture that “there’s probably a way to test it,” and offer possible answers without evidence or argument to support them (“I’m sure it depends on the material”; “that’s plus that’s minus”). Chris, in fact, remarks on not having a reason for his answer that “the plate’s positive”: “I don’t have a reason for that” (24-27).

When Vicki responds that she was also “just guessing” the opposite answer, her behavior shifts to the yellow cluster, and the rest of the group shifts with her for the remainder of the excerpt. During this part of the conversation, the students introduce ideas in the way of explanation, articulating their reasoning, but with a distinctively ambiguous tone. Chris, for example, may be serious in his feeling that the texture of the cloth would make it more likely to “hold electrons,” or he might be making a joke. Such ambiguity can be protective for a speaker who ventures an idea the group might judge unfavorably; Chris may pose the idea as a joke in order to incur less risk in sharing it. Similarly, Vicki may share Chris’s intuition as well as his nervousness about committing to it, or she may be trying to make Chris comfortable by elaborating on his suggestion (“it’s soft,” “there’s room for them”) – either seriously or in jest. Dalia chimes in by offering a reason that the plate would not hold electrons so well (“they’d slide off”), again perhaps in jest or perhaps not. Her claim that the cloth is like “toilet paper,” including a reference to a feature of popular toilet tissues (“quilted”), is more overtly humorous, but retains an association with a potentially mechanistic account of the greater electronegativity of some materials. At the end of the episode, Vicki offers to adopt Chris’s answer, perhaps out of respect for his reasoning or perhaps in a show of generosity unrelated to his ideas about the physical phenomena. Chris expresses his appreciation, possibly for her social indulgence of his off-center behavior as much as for her acceptance of his answer.

3. Epistemological framing

Group B's behavior at the outset of this snippet fits the same pattern we saw with group A, the blue cluster, and the substance of their conversation also fits the interpretation of *completing the worksheet* (or trying to) as their framing of what is taking place.

At Vicki's admission that she is "just guessing," the students' behavior and the substance of their conversation both change, apparently in response to her lighthearted tone and altered physical behavior. The students' subsequent shifting gaze and fidgeting motions correspond to their expressed awkwardness in not knowing the answers to the worksheet questions. At the same time, their giggling and smiling, along with the unrealistic content of some of their proposals, suggest that they are playing. In other words, the students communicate to each other that they frame their activity here as "joking." Still, they continue to propose and evaluate ideas: although they signal that their proposals should not necessarily be taken literally, they may also intend, on some level, to explore the ideas that they express. A framing of the activity as lighthearted can give license to play with possibilities students would be reluctant to consider in a conversation framed as "serious."

*D. Comparison across groups and tutorials*

In sum, using data from three groups of four students, we found four distinct behavioral clusters, all of which involve communication among the participants. They arose from student behavior, not from top-down categorization, and account for nearly all of the time that students spend engaged in the collaborative active-learning activities under study. The same behavioral clusters were identified for all four groups, suggesting that the four frames indicated by the behavioral clusters are common to many tutorial experiences. Tables I and II summarize the features of each behavioral cluster we identified.

1. Timelines of behavioral clusters for tutorial groups in class periods

Figure 6a depicts the timeline of group A's work on the Newton's third law tutorial, including the two episodes described above. The length of a color band represents the amount of time the group evidenced that behavioral cluster and the pie chart shows the fractions of cumulative time for each cluster during small group work. The gray band represents time in which there was no small-group interaction (*e.g.,* a demo was performed for the class). This level of depiction allows simple comparisons across groups and tutorials. For example, Figure 6b is of group B doing a tutorial on electrostatics (one episode of which is described above); Figure 6c depicts group C working in the same class as group A, on the same tutorial (Newton's third law); and 6d is of group C working on a tutorial on free-body diagrams.

As Figure 6 illustrates, different groups and different activities show different patterns of behavioral clusters in the course of a class. Group A, for example, spends about 50% of its time in the green behavioral cluster during the Newton's third law tutorial and about 30% in the blue (Figure 6a); the pattern is similar for Group B doing the electrostatics tutorial (Figure 6b). However, Group C, doing the third law tutorial, spends only 20% of its time in the green behavioral cluster and about 65% of its time in the blue (Figure 6c). Group C shows a different pattern of behavioral clusters in another tutorial (on free-body diagrams), spending about a third of the time in each of the blue and green behavioral clusters (Figure 6d). The yellow behavioral cluster accounts for only a small fraction of each group's time or, in one case, does not appear at all.

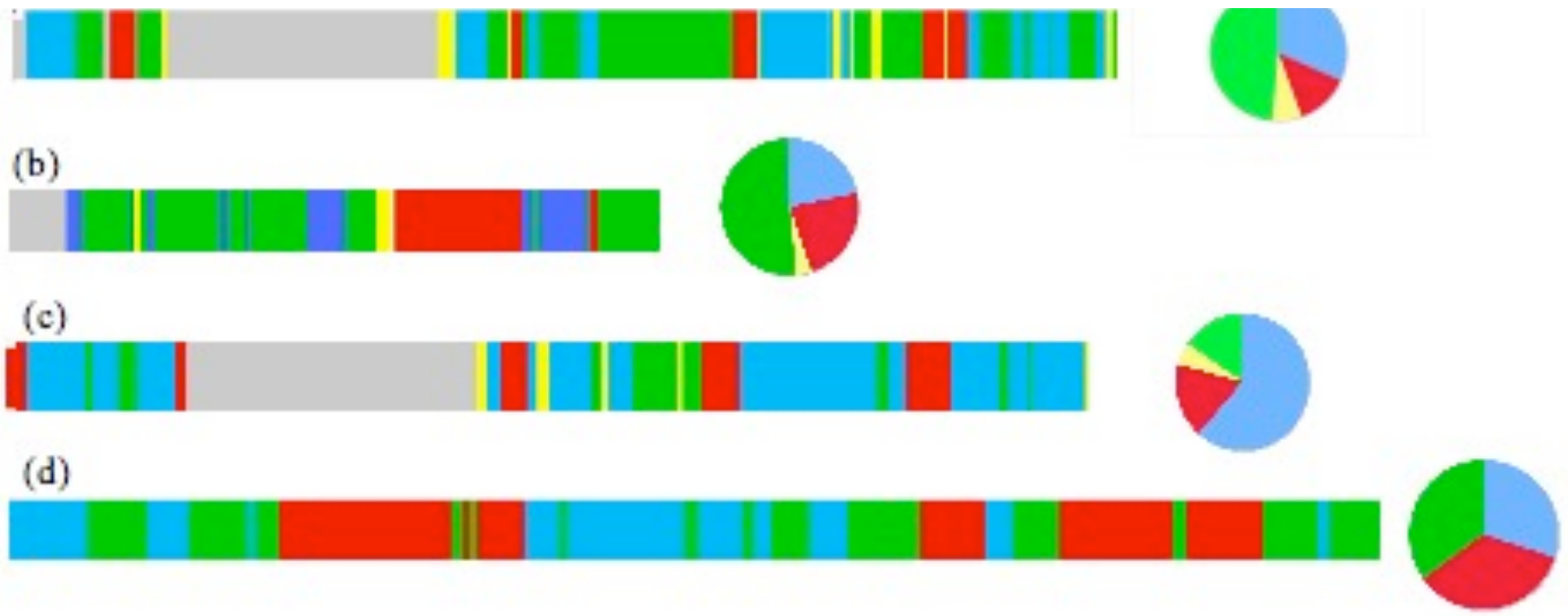

Figure 6. Timelines of behavioral clusters for tutorial groups in four different class periods, showing blue, green, red, and yellow behavioral clusters. Gray areas represent times in which there was no small-group interaction (*e.g.,* a demo was performed for the class). (a) Group A, tutorial on Newton's third law; (b) Group B, tutorial on electrostatics; (c) Group C (same classroom and TA as group A), tutorial on Newton's third law; (d) Group C, tutorial on free-body diagrams.

2. Correlation between green behavioral cluster and mechanistic reasoning

In addition to closely analyzing the interaction between student behavior and the substance of conversation, as illustrated above, we are examining the data for statistical correlations. Specifically, we have looked for correlation between the green behavioral cluster (discussion frame) and reasoning about the causal mechanisms that underlie natural phenomena. As defined by Russ, Scherr, Hammer, and Mikeska (under review), mechanistic reasoning about a physical phenomenon involves several elements: describing the target phenomenon, identifying setup conditions, identifying entities, identifying actions, identifying properties of entities, identifying the organization of entities, and "chaining" (verbally expressed logical inference). Russ et al. (under review) developed a coding scheme based on these elements as a systematic means of analyzing student conversation for evidence that they are reasoning about physical mechanism. In that scheme, the most convincing evidence of mechanistic reasoning is chaining. For example, in the first episode above, at lines 36-40, Alan says "Cause if there's equal and opposite forces, the truck... we would, if we actually saw it, we'd think the truck would hit the car and go backwards because of the force." Alan's statement contains evidence that he is reasoning about how the collision works: specifically in the categories of entities (the truck and the car), activities (the truck would hit the car, the truck would go backwards), and chaining, in the student's inference that if the forces were equal and opposite then the truck would go backwards as a result of the collision — and therefore the forces cannot be equal and opposite.[9]

In a pilot project (Conlin, Gupta, Scherr, & Hammer, 2007), twenty-minute videoclips from six tutorial sessions chosen at random were coded for behavioral clusters (without transcript, as described above). In a separate analysis, the sessions were transcribed and coded for (1) instances of some level of mechanistic reasoning and (2) instances of chaining. Two independent coders agreed on 90% of the behavior codes and 87% of the mechanistic reasoning and chaining codes before discussion. The codes for behavior, mechanistic reasoning, and chaining were then matched for each five seconds of the tutorial session. Looking across groups, 81% of the chaining occurred during the discussion frame (green behavioral cluster). This result further validates our observation that the discussion frame consists primarily of substantive conversation about physics ideas.

Our conjecture is that this correlation reflects a consonance between students' framing their activity as discussion and their reasoning about causal mechanisms, a consonance that may occur for several reasons. First, gesturing can help individuals reason about mechanisms, by performing simulations or depictions of the entities and their organization and activity; second and similarly, they provide a means for communicating those ideas to others (Scherr, in press). For these reasons, a conversational mode that affords gesturing may be more conducive to mechanistic reasoning, and students interested to discuss mechanism may be more inclined to enter such a mode. Another possibility is that the correlations we find between the discussion frame and mechanistic reasoning may reflect more simply the fact that to engage in chaining requires more extended conversational turns and more extended attention by conversational partners. In any case, behaviors in the green cluster are associated with scientifically valuable reasoning and interactions, and the absence of those behaviors may be cause for concern.

## VI. SUMMARY AND QUESTIONS FOR FUTURE RESEARCH

The behavioral clusters identified in the above analysis are evidence of students' epistemological framing. The behavioral cluster labeled "green," for example, which includes animated speech, eye contact, and gesturing, indicates a framing of the activity as discussing one another's conceptual ideas. Those behaviors not only display that type of discussion – they also promote it. The association between framing and behavior advances the possibilities for identifying students' framing of collaborative active-learning activities. The methodology presented is powerful, reliable, and efficient, and adds substantially to previous analysis methods based on student speech acts.

Resource-based views of epistemological framing encourage theoretical continuity between individual and group (*i.e.,* cognitive and discourse) dynamics. A frame is a locally coherent pattern of activations that emerges from the interactions of many cognitive elements; these elements may be within an individual's mind or across multiple individuals, and the coherences may be within an individual's reasoning or among members of a group. Our approach illustrates an analyzable, codable dynamic between the way people behave in groups and the substance of their thinking. For example, a student who makes a kinesthetically evocative gesture may be at once using an intuitive sense of mechanism and, at the same time, communicating a metamessage (Tannen, 1993) about what sort of conversation is taking place. Another student may move from blue to green behaviors, stimulating others to follow her lead (or not). To the extent that our account connects students' behavior, conceptual reasoning, and epistemological framing, it integrates individual cognition with group dynamics, and makes visible the interplay between conceptual reasoning and views of knowledge and learning at the level of both the individual and the group.

This analysis methodology makes a range of research questions empirically accessible. The examples above suggest many such questions: What frames occur in various classroom activities? In which frames do certain desirable activities (including cognitive activities) occur? What precipitates shifts into (or out of) desirable frames? Do particular learning activities have characteristic frames associated with them? Do student groups have characteristic framings of particular activities that they undertake together? Does a group's framing of particular activities change over the course of an hour, or a term? How might we best promote the substantive student discussion that we believe is essential to high-quality physics instruction? How can we account for the observed deterioration in student expectations and epistemologies?

The question of how students frame tutorials does not appear to have a simple answer. The methodology we have described focuses on framing and shifts of framing in the immediate moment, and it provides evidence of both variability and local coherences. Within tutorial sessions, we are developing case studies of the particular dynamics by which a group moves into or out of framing the immediate activity as a discussion. These case studies may shed light on the reasons for the correlations we noted between the discussion frame and mechanistic reasoning. They may also shed light on how students' framing of tutorials as a whole influences their moment-to-moment activity, *e.g.* does the fact that group C spends so little time in discussion reflect the students' framing the tutorial session more as completing a task than as sensemaking? Ultimately, we are interested in framing at the larger scale: that is, in students' sense of "what is it that's going on here" in the tutorial as a whole. The timelines presented in Figure 6 present some evidence of overall patterns: some groups spend the majority of some tutorials in animated discussion, while for other groups and tutorials, the greater part of the time is spent concentrating on the worksheet. Looking across sessions, we are studying how patterns of behavioral clusters may vary for a particular group over the course of a term (*e.g.,* does group C come to spend more time in discussions during tutorial sessions later in the semester?) or whether the patterns vary by tutorial (*e.g.,* does one version of the third law tutorial lead to more student discussion?) But these timelines clearly provide only a limited view, and "time on task" may not indicate the importance of a particular framing to the experience of tutorials in general. Some activities might have, for example, a short duration but high intensity, in the sense of having great importance for the participants.

## ACKNOWLEDGEMENTS

We are grateful to Andrew Elby, E. F. Redish, and the other members of the Physics Education Research Group at the University of Maryland for substantive discussions of this research. Raymond Hodges, Ayush Gupta, and Luke Conlin served as additional coders for inter-rater reliability. This work was supported in part by the National Science Foundation (REC 0440113).

## ENDNOTES

[1] Frames, scripts, and schemata are related and overlapping terms in the fields of linguistics, artificial intelligence, cognitive psychology, social psychology, sociology, anthropology, and other disciplines. An overview and history of the uses of these related terms appears in Ch. 1 of Tannen (1993).

[2] Linguistic markers include omissions, repetitions, negatives, modals, *etc.* and are described in Ch. 1 of Tannen (1993).

[3] There is the possibility of the speaker acting as though she is engaged in a lively discussion while in fact not experiencing genuine engagement. For frame analysis in the context of acting, see Goffman (1986).

[4] Students' names are pseudonyms.

[5] This turns out to be the case with epistemological state measured by pre-post MPEX and conceptual state measured by fractional gains on the FCI. Strong gains were obtained in both measures. These results will be documented elsewhere.

[6] These groups were not the only "watchable" groups; some videotaped groups were not classified. These groups included average, above-average, and below-average students as measured by their overall performance in the course. "Watchable" groups were not typically better than average or more successful in mastering the conceptual content of the tutorials, except to the extent that on-task groups tend to do better than off-task groups.

[7] This result is partially to be expected: since peers' activity together is mutually constructed, it is natural that they should share their framing of a given situation and therefore participate in the same behavioral cluster. On the other hand, it seems possible that individuals could "opt out" of the framing mutually constructed by the others in the group. This paper reports brief occurrences of such opting-out or mismatch of framings. However, the great majority of the time, peers share the same behavioral cluster.

[8] Sometimes, for example, they exhibit "green" behaviors with a TA nearby or even joining the discussion.

[9] Her reasoning is not correct; see Russ & Hutchison (2006) for a discussion about correctness and mechanistic reasoning.